\documentclass[doublecol]{epl2}
\usepackage[latin1]{inputenc}
\usepackage{graphicx}
\usepackage{dcolumn}
\usepackage{bm}
\usepackage{amsmath}

\newcommand{\si}[1]{\text{sgn}(}

\title{Complete renormalization group calculation up to two-loop order of an effective two-band model for iron-based superconductors}
\shorttitle{Complete RG calculation up to two-loop order of an effective two-band model for iron-based superconductors} 

\author{Vanuildo S. de Carvalho\inst{1} \and Hermann Freire\inst{1}\footnote{Corresponding author: hermann@if.ufg.br}}

\institute{\inst{1} Instituto de Física, Universidade Federal de Goiás, 74.001-970, Goiânia-GO, Brazil}


\abstract{We perform a renormalization group (RG) study up to two-loop order of an effective low-energy two-band model to describe some of the recently discovered iron-based superconductors. Our starting point is the itinerant electronic model proposed by Chubukov \emph{et al.} [Phys. Rev. B \textbf{78}, 134512 (2008)], which displays two small, almost nested Fermi pockets with one hole pocket centered at $(0,0)$ and one electron pocket centered at $\mathbf{Q} = (\pi,\pi)$ in
the folded Brillouin zone. We then proceed to implement a complete two-loop RG calculation for this model
of four-point vertex corrections, quasiparticle weight and several order-parameter susceptibilities in order to evaluate the robustness of one-loop RG results available in the literature with respect to including self-energy effects and higher-order quantum fluctuations.}

\begin{document}

\maketitle

\textbf{Introduction.} -- The recent experimental observation \cite{Hosono,Hosono2} that the iron-based pnictides (such as LaFeAsO and SrFe$_2$As$_2$, to name but a few) exhibit unconventional superconductivity at critical temperatures up to $T_c=55K$ sparked big efforts in the community of strongly correlated systems to understand these materials. This is because they are the first high-$T_{c}$ superconductors ever discovered which are outside the well-known copper-oxide family. Since electron-phonon coupling in these iron compounds appears to be very small \cite{Boeri}, a purely electronic mechanism for superconductivity emerges as a strong possibility. Indeed, in many aspects, the pnictides resemble the physics of the cuprate superconductors: these two classes of compounds are effectively two-dimensional materials which exhibit antiferromagnetism \cite{Cruz} at zero doping and both become a superconductor upon doping. This observation could imply that the discovery of the underlying mechanism of superconductivity in the iron pnictides might also give important insight into solving the longstanding cuprate high-$T_{c}$ superconductivity problem.

By contrast, there are some crucial differences between the pnictides and the cuprates, which makes the former materials also very interesting from a fundamental point of view. Unlike the cuprates which always display a localized Mott insulating phase at low doping, the pnictides are instead semi-metals for this doping regime \cite{Cruz}. Besides, there is growing consensus in the community that some of these iron-based materials remain itinerant for all doping levels \cite{Norman}. This suggests that the pnictides are in fact less correlated than the cuprates and, therefore, weak coupling theories could be a valid starting point to describe at least qualitatively the properties of these former materials \cite{Yang,Anisimov}. Moreover, from angle-resolved photoemission spectroscopy (ARPES) experiments \cite{Liu,Kondo} and theoretical band-structure considerations \cite{Singh,Mazin}, it is now generally accepted that the low-energy electronic structure of these materials are in some respects similar to one another, with a Fermi surface consisting essentially of two small electron pockets centered around $M=(\pi,\pi)$-point and two (or three) small hole pockets centered around $\Gamma=(0,0)$-point in the so-called folded Brillouin zone.

Motivated by these experimental results, many researchers have put forward various types of multi-band electronic models aiming to explain some of the key properties displayed by these materials, most notably, their phase diagram and also the symmetry of the corresponding superconducting gap. These studies included two-band \cite{Chubukov}, four-band \cite{Korshunov,Tesanovic,Honerkamp,Thomale} and five-band models \cite{Lee}. These models typically contain several competing ordering tendencies at low-energies, which must be in principle treated on equal footing. In this respect, a promising theoretical framework which is tailored for describing such systems turns out to be the renormalization group (RG) approach in view of its unbiased nature \cite{Shankar}.

\begin{figure}[t]
  \centering
  \includegraphics[height=6cm]{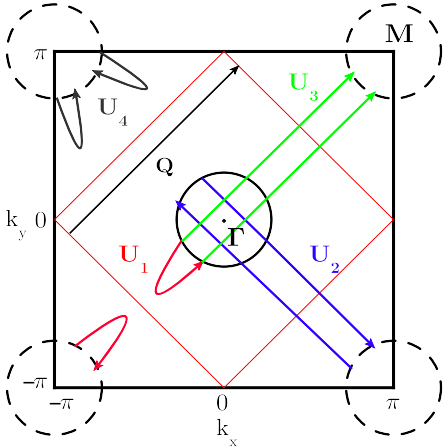}\\
  \caption{(Color online) The Fermi pockets of the two-band model analyzed in this work. We also show schematically the interaction
  processes included in this model: $U_{1}$ and $U_{2}$ stand for, respectively, interband couplings with (0,0) and $\mathbf{Q}=(\pi,\pi)$ momentum transfer, $U_{3}$ corresponds to interband pair scattering, and $U_{4}$ refers to all intraband interactions.}\label{FS}
\end{figure}

Soon after the discovery of the pnictides, Chubukov \emph{et al.} proposed a simple but ingenious effective two-band model \cite{Chubukov}, which displays two small, almost nested Fermi pockets in the folded Brillouin zone to describe such materials. By applying a one-loop RG approach -- which is essentially equivalent to summing the so-called parquet diagrams up to infinite order -- they concluded that at zero doping the model displays a antiferromagnetic spin-density wave (SDW) phase in agreement with experiments and, upon doping, a superconducting (SC) phase emerges with a extended $s^{\pm}$-wave gap symmetry, as first suggested in Ref. \cite{Mazin}. In another important work \cite{Lee} done by Wang \emph{et al.}, a more complex five-band model for the iron pnictides was analyzed using a functional generalization of the one-loop RG approach, and their results also point to an antiferromagnetic phase for the undoped system and a robust $s^{\pm}$-wave pairing state which arises at larger doping. In order to further demonstrate that these results did not depend critically on some details of the band-structure of the above models and the approximations used, the authors in Ref. \cite{Honerkamp} set out to discuss a four-band model within the one-loop functional RG treatment and their data provided yet another confirmation of the previous RG results regarding SDW and $s^{\pm}$-wave symmetry superconductivity as leading instabilities obtained within the two-band and five-band models.

Another aspect of these iron-based superconductors concerns their gap structure -- \emph{i.e.} the $\mathbf{k}$-dependent variation of the SC order parameter within a given symmetry class -- and the related question of existence (or absence) of nodes on them. Two works recently addressed this issue within a one-loop RG approach by comparing the results obtained within a four-band and a five-band model \cite{Thomale2,Chubukov2}. As a result, Thomale \emph{et al.} concluded that in the four-band case there should appear nodal gaps on the electron pockets, while the hole pockets are always nodeless. By contrast, if an additional hole pocket is included in the model (hence, resulting in a five-band model), the net effect of this would be turning the nodal gap on the electron pockets nodeless. This could explain why some compounds seem to exhibit experimentally nodal gaps and others not. This conclusion is shared by the authors in Ref. \cite{Chubukov2} who also gave a thorough analysis of the two-band model case. In this latter work, it was shown that the two-band model naturally describes nodeless gaps on both electron and hole pockets which agrees qualitatively with the five-band scenario. This could suggest that both two-band and five-band models might be in the same universality class and therefore have the same low-energy physics. Consequently, this would imply that the two-band model might be a good, minimal low-energy effective model in order to describe some iron-based superconductors which display nodeless gaps on all pockets.

For this reason, we revisit in this work the two-band model discussed previously by Chubukov \emph{et al.} within a one-loop RG approach. As a result, we report here a complete two-loop RG calculation for this model of all vertex corrections, self-energy and several order-parameter susceptibilities with the important goal of evaluating the robustness of one-loop RG results available in the literature with respect to including self-energy effects and higher-order quantum fluctuations.

\begin{figure}[t]
  \centering
  \includegraphics[height=1.5cm]{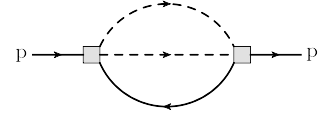}\\
  \caption{Sunset diagram for the self-energy at two-loop order which yields a non-analytic contribution for energies larger than the Fermi energy in the present two-band model.}\label{Selfenergy}
\end{figure}

\textbf{Model.} -- Our starting point is the symmetric phase of the two-band model proposed in Ref. \cite{Chubukov} that displays two small, almost nested Fermi pockets -- with one hole pocket centered at $\Gamma$-point and one electron pocket centered at $M$-point in the folded Brillouin zone (see Fig. 1) -- and which includes also both interorbital and intraorbital interactions. If we use a coherent-state functional integral representation of the resulting Hamiltonian after a suitable canonical transformation (for a thorough discussion on this point, see Ref. \cite{Chubukov}), the model at $T = 0$ and constant chemical potential $\mu=E_F$ becomes described by the action $S=S_0 + S_{int}$, where

\vspace{-0.2cm}

\begin{eqnarray}
S_{0}=\sum_{\mathbf{p},\sigma}[(-i\omega+\epsilon_{\mathbf{p}}^{c})\overline{c}_{\mathbf{p}\sigma}c_{\mathbf{p}\sigma}
+(-i\omega+\epsilon_{\mathbf{p}}^{f})\overline{f}_{\mathbf{p}\sigma}f_{\mathbf{p}\sigma}], \label{action1}
\end{eqnarray}

\noindent and

\vspace{-0.2cm}

\begin{align}\label{action2}
S_{int}&=\,U^{(0)}_{1}\sum_{\substack{{\mathbf{p_1,p_2,p_3}} \\ {\sigma,\sigma'}}}
\overline{c}_{\mathbf{p_4}\sigma}\overline{f}_{\mathbf{p_3}\sigma'}f_{\mathbf{p_2}\sigma'}c_{\mathbf{p_1}\sigma}\nonumber\\
&+\, U^{(0)}_{2}\sum_{\substack{\mathbf{p_1,p_2,p_3} \\ \sigma,\sigma'}}\overline{f}_{\mathbf{p_4}\sigma}\overline{c}_{\mathbf{p_3}\sigma'}f_{\mathbf{p_2}\sigma'}c_{\mathbf{p_1}\sigma}\nonumber\\
&+\,\frac{U_{3}^{(0)}}{2}\sum_{\substack{\mathbf{p_1,p_2,p_3} \\ \sigma,\sigma'}}(\overline{f}_{\mathbf{p_4}\sigma}\overline{f}_{\mathbf{p_3}\sigma'}c_{\mathbf{p_2}\sigma'}c_{\mathbf{p_1}\sigma}\nonumber\\
&+\overline{c}_{\mathbf{p_4}\sigma}\overline{c}_{\mathbf{p_3}\sigma'}f_{\mathbf{p_2}\sigma'}f_{\mathbf{p_1}\sigma})\nonumber\\
&+\,\frac{U_{4}^{(0)}}{2}\sum_{\substack{\mathbf{p_1,p_2,p_3} \\ \sigma,\sigma'}} (\overline{f}_{\mathbf{p_4}\sigma}\overline{f}_{\mathbf{p_3}\sigma'}f_{\mathbf{p_2}\sigma'}f_{\mathbf{p_1}\sigma}\nonumber\\
&+\overline{c}_{\mathbf{p_4}\sigma}\overline{c}_{\mathbf{p_3}\sigma'}c_{\mathbf{p_2}\sigma'}c_{\mathbf{p_1}\sigma}),
\end{align}

\noindent where $\mathbf{p_4}=\mathbf{p_1}+\mathbf{p_2}-\mathbf{p_3}$ and the volume $V$ has been set equal to unity. We assume here, for simplicity, that $\epsilon_{\mathbf{p}}^{c}=E_F - \mathbf{p}^2/2m$ and $\epsilon_{\mathbf{p}+\textbf{Q}}^{f}=-\epsilon_{\mathbf{p}}^{c}$. Besides, $\overline{c}_{\mathbf{p}\sigma}$ and $c_{\mathbf{p}\sigma}$ are, respectively, the creation and annihilation Grassmann
fields for fermions with spin projection $\sigma$ that are near $\mathbf{k}=(0,0)$, $\overline{f}_{\mathbf{p}\sigma}$ and
$f_{\mathbf{p}\sigma}$ are the creation and annihilation Grassmann fields for fermions with spin projection $\sigma$ that are close to $\mathbf{\mathbf{Q}}=(\pi,\pi)$, and $U^{(0)}_{i}$ (for $i=1..4$) are the microscopic (bare) coupling constants of the model which are displayed schematically in Fig. 1. The above action defines our bare quantum field theory which must be regularized in the ultraviolet by restricting the energies to $|\omega|\leq \Lambda_0$ (\emph{i.e.}, $W=2\Lambda_0$ represents the bandwidth of the model).

\begin{figure}[t]
  \includegraphics[height=3cm]{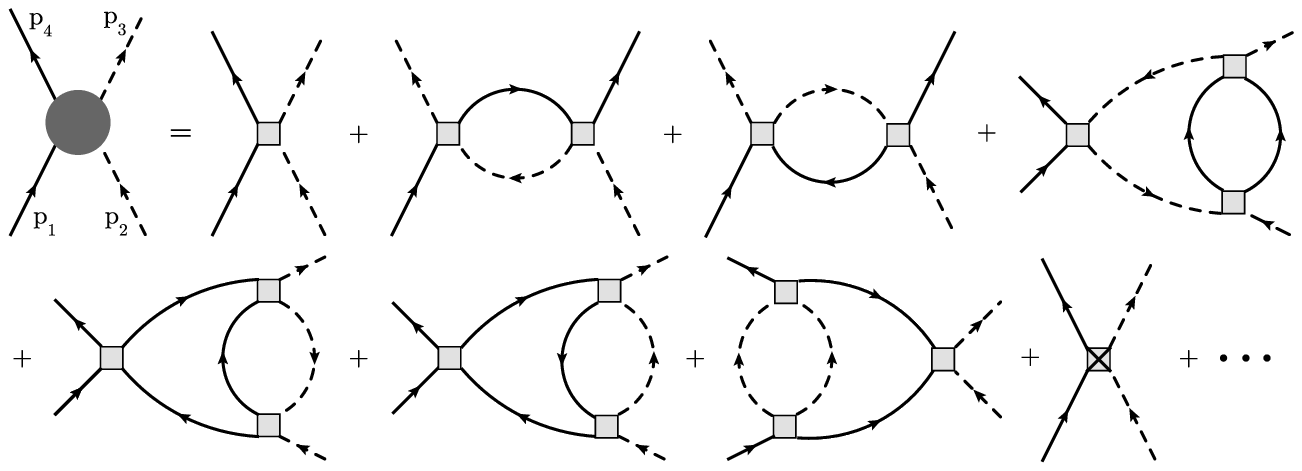}
  \includegraphics[height=3cm]{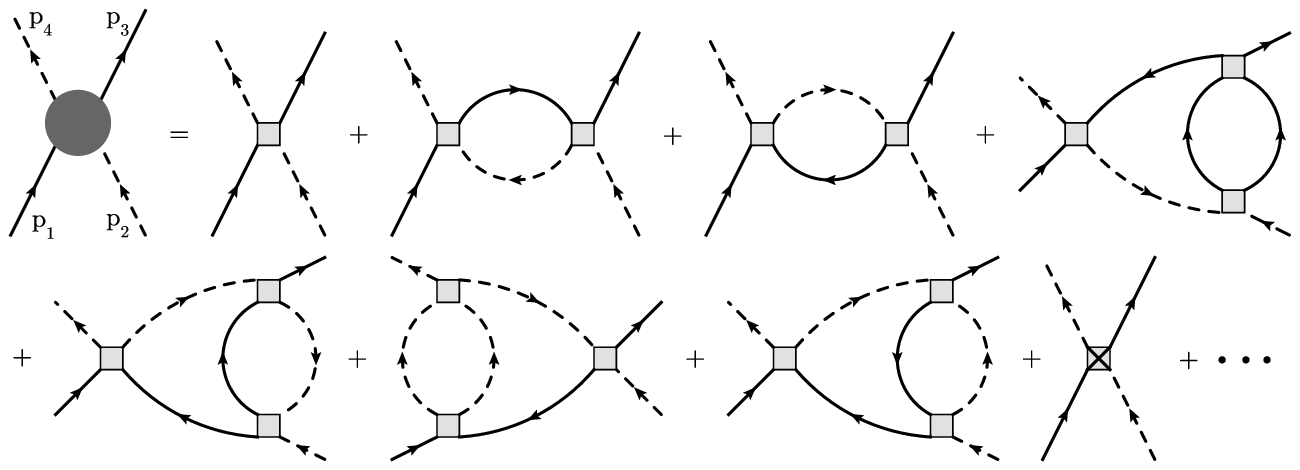}
  \includegraphics[height=1.5cm]{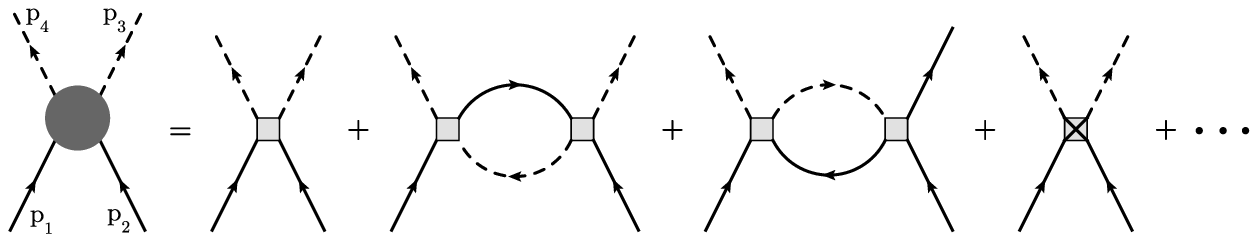}
  \includegraphics[height=3.5cm]{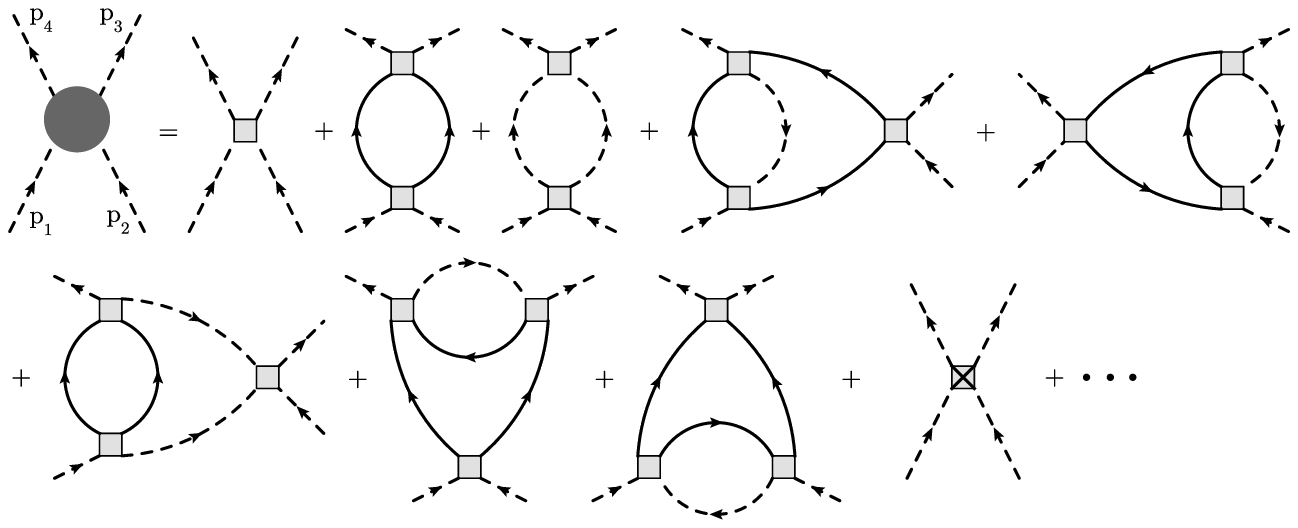}
  \caption{Schematic representation of the Feynman diagrams of vertex corrections included in the present two-loop RG calculation. The single-particle propagators are represented by either solid or dashed lines according to their association with the corresponding Fermi pockets. The diagrams with crossed squares represent the counterterms.}\label{Couplings}
\end{figure}

Since both hole and electron pockets of the iron pnictides are small compared to the bandwidth $W$ as measured from ARPES experiments, we shall concentrate throughout this work only on the physical regime of the two-band model in which the energies are actually larger than the Fermi energy ({$|\omega|> E_F$}). In this way, if we calculate the self-energy of the model up to two-loop order for this regime, we obtain that its non-analytic contribution (Fig. 2) is given {approximately} by

\vspace{-0.4cm}

\begin{eqnarray}
\Sigma(i\omega,{\mathbf{p}=\mathbf{k}_F})\approx\frac{[U_3^{(0)}]^{2}}{4} N^2(0) i\omega\ln\left(\frac{\Lambda_0}{i\omega}\right)+...,
\end{eqnarray}

\noindent with {$\mathbf{k}_F$ being the Fermi vector} and ${N(0)}=m/2\pi$ is the constant density of states of a two-dimensional Fermi gas. The presence of this non-analyticity
in the self-energy is a generic feature of this model and several logarithmic divergences also appear if one calculates four-point vertex corrections and order parameter susceptibilities within perturbation theory up to two-loop order.

\textbf{Method.} -- The field-theoretical RG approach up to two loops that we shall use now is standard \cite{Peskin} and details also appear elsewhere in the context of another fermionic model \cite{Freire}. In order to circumvent the problem of logarithmic singularities and non-analyticities emerging in the low-energy limit of the present model within perturbation theory (see Fig. 3), the field-theoretical RG strategy consists of rewriting the unobserved bare quantities of the microscopic model defined by Eqs. (\ref{action1}) and (\ref{action2}) in terms of the experimentally observed renormalized parameters plus appropriate counterterms. These counterterms have the main effect of regularizing classes of diagrams to a given order in the model at a floating new energy scale $\Lambda$, such that the renormalized perturbation theory becomes well-defined in the low-energy limit (\emph{i.e.} $\Lambda\rightarrow 0$). An important point we wish to stress here is that both coupling constants and fermionic fields of the model must be renormalized at two-loop RG level. Therefore, we must define: $c_{\mathbf{p}\sigma}=Z^{1/2}c_{\mathbf{p}\sigma}^{R}$, $f_{\mathbf{p}\sigma}=Z^{1/2}f_{\mathbf{p}\sigma}^{R}$, and $U^{(0)}_{i}=N^{-1}(0)Z^{-2}(u_{i}+\Delta u_{i})$, where $Z=(1-\partial \Sigma(i\omega,{\mathbf{p}=\mathbf{k}_F})/\partial(i\omega) |_{\omega=0})^{-1}$ is the quasiparticle weight, $c_{\mathbf{p}\sigma}^{R}$ and $f_{\mathbf{p}\sigma}^{R}$ stand for the renormalized fields, $u_{i}$ (for $i=1..4$) represent the corresponding dimensionless renomalized coupling constants of the model, and $\Delta u_{i}$ refer to the counterterms which must be calculated order by order in perturbation theory. Since this program is successfully accomplished, the field theory model analyzed here is indeed renormalizable.

\textbf{Results.} -- We can now adjust the counterterms $\Delta u_{i}$ such that all divergences
are exactly canceled in our series expansion up to two-loop order. But the price we pay for
this is the appearance of a new scale $\Lambda$ with all physical
quantities now depending on this scale. By contrast,
the original model has no information about this quantity, \emph{i.e.} the bare parameters do not depend on $\Lambda$.
This leads us to the renormalization group conditions
for the bare couplings of the model, \emph{i.e.} $\Lambda(dU_{i}^{(0)}/d\Lambda)=0$. As a result, the RG flow equations for the renormalized dimensionless couplings
at two-loop order become

\vspace{-0.3cm}

\begin{eqnarray}
\dot{u}_{1}=&+&(u_{1}^{2}+u_{3}^{2})+(u_{1}-u_{2})u_{3}^2\nonumber\\
&-&2u_{4}(u_{1}^2 + u_{2}^2 - u_{1}u_{2}+ \frac{1}{2}u_{3}^2),\\
\dot{u}_{2}=&+&2(u_{1}-u_{2})(u_{2}+u_{2}u_{4})-2u_{2}u_{3}^2,\\
\dot{u}_{3}=&+&2u_{3}(2u_{1}-u_{2}-u_{4})-u_{3}^{3},\\
\dot{u}_{4}=&-&(u_{3}^{2}+u_{4}^{2})-u_{4}u_{3}^2-2u_{2}^2 u_{3}-2u_{3}^3 \nonumber\\
&+&2u_{1}u_{3}^2+3u_{2}u_{3}^2-2u_{1}^3
-2u_{1}u_{2}^2+2u_{1}^2 u_{2},\label{u4}
\end{eqnarray}

\noindent where the derivatives are taken with respect to $\xi=(1/2)\ln(\Lambda_0/\Lambda)$. The initial conditions for this system of differential equations are naturally given by the microscopic interactions, \emph{i.e.} $u_i (\xi=0)=N(0)\, U_{i}^{(0)}$ (for $i=1..4$). A schematic representation of the Feynman diagrams corresponding to the vertex corrections up to two-loop order are displayed in Fig. 3.

First, we focus our attention on the numerical solution of the flow equations for the renormalized couplings as a
function of $\Lambda$. This was performed by means of fourth-order Runge-Kutta method.
We analyze simultaneously both one-loop and two-loop RG flows for a direct comparison of the two approaches. In both cases, even though all couplings indeed diverge at low energies ($\xi\rightarrow \infty$), their ratio approach infrared (IR) stable fixed points in this limit (Fig. 4). Most importantly, as can be clearly see
from this figure, the inclusion of two-loop order fluctuations in the RG scheme has the effect of changing the fixed point structure of the field theory model.

In the one-loop RG approach, as first obtained in Ref. \cite{Chubukov}, there are two IR stable fixed points given by $(u_2/u_1)\rightarrow 0$, $u_3=\pm\sqrt{5}u_1$, $u_4=-u_1$, which control the low-energy dynamics of the model. Therefore, from the RG analysis of the corresponding susceptibilities up to one loop, the authors in Ref. \cite{Chubukov} concluded that the leading coupling in this limit turns out to be the $u_3$ pair scattering which appears to be the main responsible for inducing a SDW instability at zero doping and which gives rise to an extended $s^{\pm}$-wave superconducting instability at larger doping in the model.

\begin{figure}[t]
  \includegraphics[height=5.3cm]{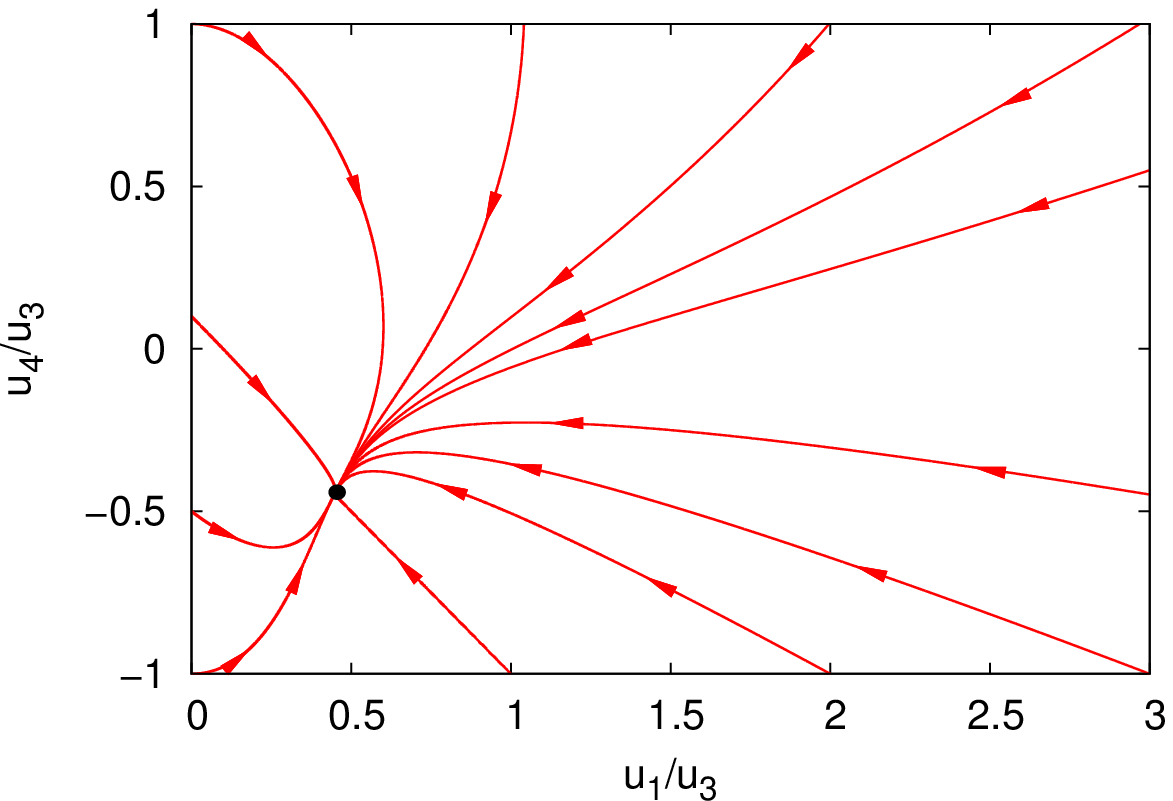}
  \includegraphics[height=5.3cm]{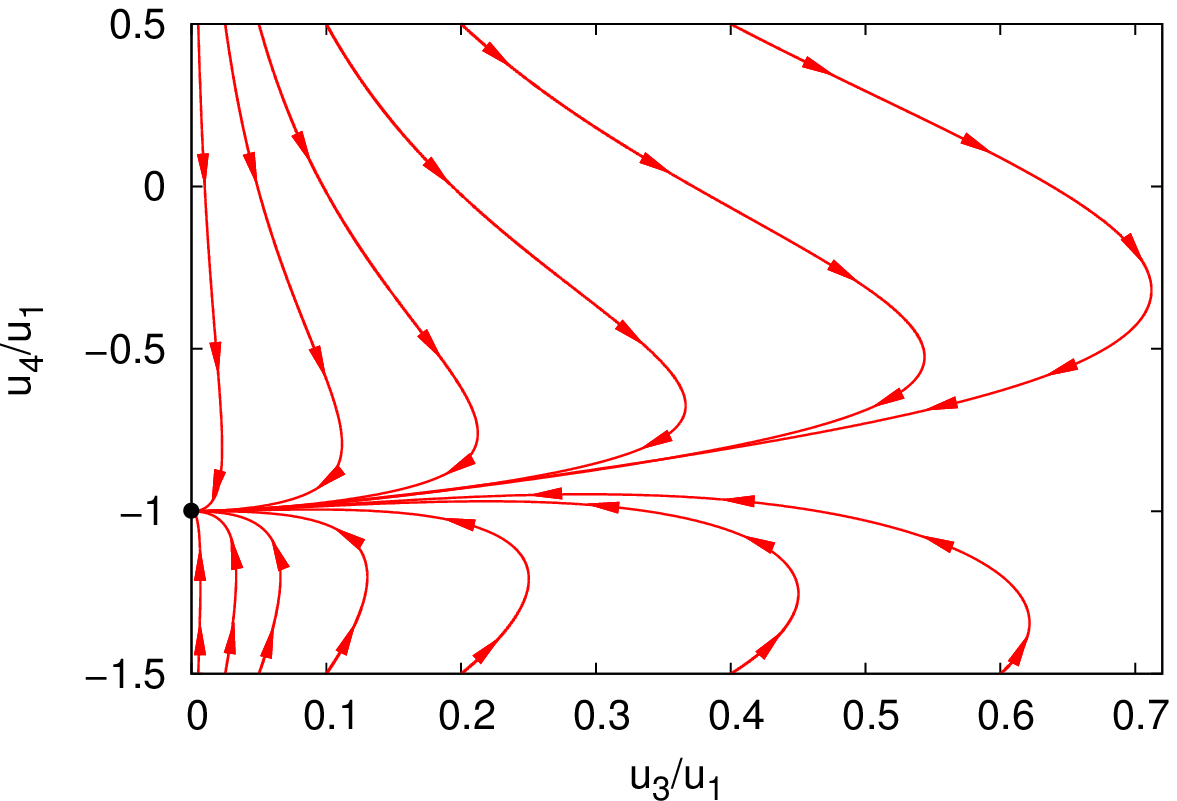}
  \caption{(Color online) RG flow at one-loop order (upper panel) and two-loop order (lower panel). {(Note that the axes in the above plots are different.)}}\label{Diagrams}
\end{figure}

On the other hand, in the complete two-loop order RG approach implemented here in this work, there is only a single IR stable fixed point which is given instead by $(u_2/u_1)\rightarrow 0$, $(u_3/u_1)\rightarrow 0$, $u_4=-u_1$. Hence, higher-order fluctuations shift the dominance
of the coupling constants. Although the $u_3$ coupling still diverges in the low-energy limit, it does so at a slower rate compared to the one-loop RG approach. As a consequence, the interaction processes given by the interband and intraband forward scatterings (\emph{i.e.} $u_1$ and $u_4$, respectively) eventually overcome the $u_3$ pair interaction in this limit. It is interesting to note that the new IR fixed point found here at two-loop RG level still satisfies the $SO(6)$ symmetry condition \cite{Podolsky}. Moreover, the quasiparticle weight $Z$ renormalizes very weakly in this regime and always remains closer to unity, thereby indicating Fermi liquid behavior.

\begin{figure}[t]
  \includegraphics[height=1.95cm]{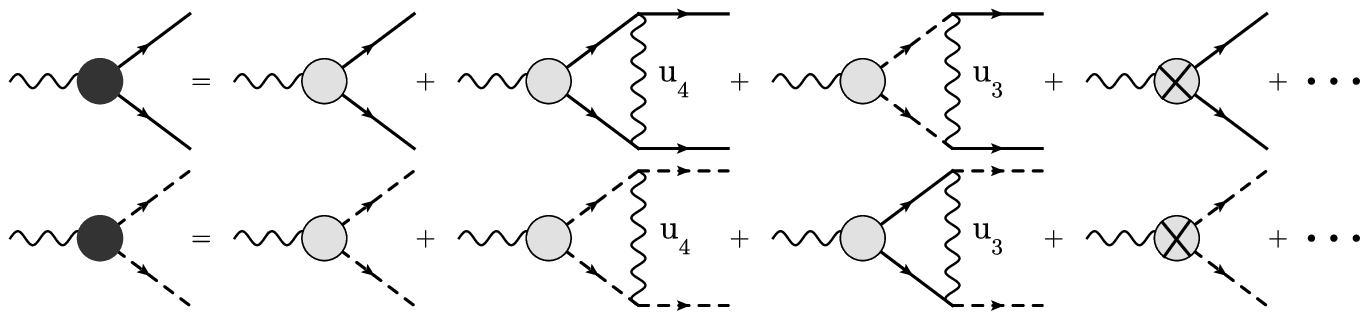}
  \includegraphics[height=1.95cm]{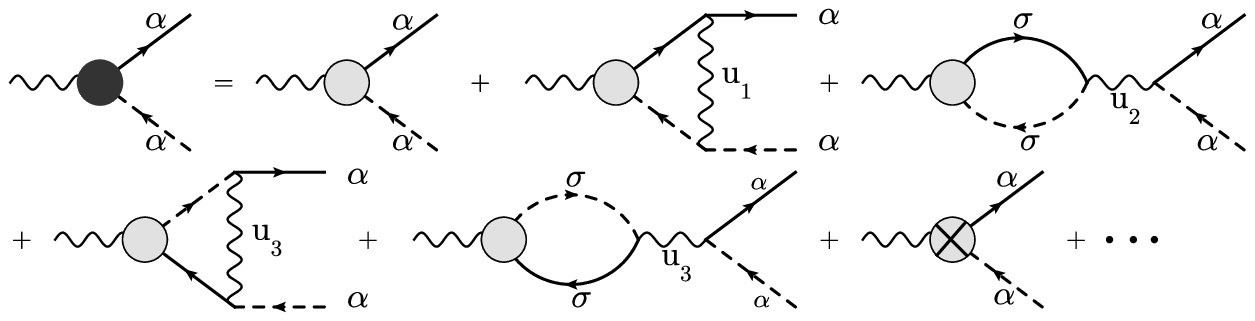}
  \caption{Feynman diagrams for the response vertices for both superconducting [$\mathcal{T}_{SC,c(f)}^{(0)}$] and density-wave [$\mathcal{T}_{DW}^{(0)\,\alpha\alpha}$] orders. The diagrams with crosses represent the corresponding counterterms.}\label{Diagrams}
\end{figure}

In order to verify if the interband and intraband forward interactions are able to induce density-wave and pairing instabilities in the model, it is important to calculate the corresponding susceptibilities by introducing an infinitesimal external field in the appropriate channel and evaluating its linear response. Therefore, we must add to the action that describes the present model the following term

\vspace{-0.4cm}

\begin{eqnarray}
S_{ext}&=&\sum_{\mathbf{k,\sigma}}\mathcal{T}_{SC,c}^{(0)}\,\overline{c}_{\mathbf{k},\sigma}\overline{c}_{\mathbf{-k},-\sigma}
+\sum_{\mathbf{k,\sigma}}\mathcal{T}_{SC,f}^{(0)}\,\overline{f}_{\mathbf{k},\sigma}\overline{f}_{\mathbf{-k},-\sigma}\nonumber\\
&+& \sum_{\mathbf{k,\alpha,\beta}}\mathcal{T}_{DW}^{(0)\,\alpha\beta}\,\overline{c}_{\mathbf{k},\alpha}f_{\mathbf{k+Q},\beta}
+h.c.,
\end{eqnarray}

\noindent where $\mathcal{T}_{SC,c(f)}^{(0)}$ and $\mathcal{T}_{DW}^{(0)\,\alpha\beta}$ are the bare response vertices for the superconducting and density-wave orders, respectively.
This added term will generate new Feynman diagrams
-- the three-legged vertices displayed in Fig. 5 -- which will
also generate new logarithmic singularities in the low-energy
limit of our field theory model. Therefore, we must regularize these divergences (see, \emph{e.g.}, Ref. \cite{Freire2} in the context of another fermionic model) by defining the renormalized response vertices and the corresponding counterterms
as follows: $\mathcal{T}_{SC,c(f)}^{(0)}=Z^{-1}(\mathcal{T}_{SC,c(f)}+\Delta \mathcal{T}_{SC,c(f)})$ and
$\mathcal{T}_{DW}^{(0)\,\alpha\beta}=Z^{-1}(\mathcal{T}_{DW}^{\alpha\beta}+\Delta \mathcal{T}_{DW}^{\alpha\beta})$. Again, by invoking the RG condition for the bare quantities of the model $\Lambda(d\mathcal{T}_{i}^{(0)}/d\Lambda)=0$
(for $i=SC$ and $DW)$, we obtain the RG flow equations for the response vertices

\vspace{-0.4cm}

\begin{eqnarray}
&&\hspace{-0.7cm}\dot{\mathcal{T}}_{SC,c(f)}=-\left(u_4+\frac{u_{3}^{2}}{2}\right)\mathcal{T}_{SC,c(f)}
-u_3\mathcal{T}_{SC,f(c)},\\
&&\hspace{-0.7cm}\dot{\mathcal{T}}_{DW}^{\alpha\alpha}=\left(u_1-\frac{u_{3}^{2}}{2}\right)\mathcal{T}_{DW}^{\alpha\alpha}-u_2\sum_{\sigma=\alpha,\beta}
\mathcal{T}_{DW}^{\sigma\sigma}\nonumber\\&&\hspace{+0.5cm}-u_3\,\delta_{\alpha,-\beta}[\mathcal{T}_{DW}^{\beta\beta}]^*,
\end{eqnarray}

\noindent with $\alpha,\beta=\uparrow,\downarrow$, and the derivatives here are also taken with respect to $\xi$. By symmetrizing these response vertices, we obtain
the following order parameters

\vspace{+0.3cm}

$\left\{%
\begin{array}{ll}
    \mathcal{T}_{SC}^{(s^{\pm})}=\mathcal{T}_{SC,c}-\mathcal{T}_{SC,f},\\
    \mathcal{T}_{SC}^{(s)}=\mathcal{T}_{SC,c}+\mathcal{T}_{SC,f},\\
    \mathcal{T}_{CDW(SDW)}=\mathcal{T}_{DW}^{\uparrow\uparrow}\pm\mathcal{T}_{DW}^{\downarrow\downarrow},\\
    \mathcal{T}_{CDW(SDW)\pm}=\mathcal{T}_{CDW(SDW)}\pm\mathcal{T}^{*}_{CDW(SDW)},
\end{array}%
\right.$

\vspace{+0.3cm}

\noindent where $\mathcal{T}_{SC}^{(s^{\pm})}$ and $\mathcal{T}_{SC}^{(s)}$ represent, respectively, extended s-wave and conventional s-wave superconducting orders, $\mathcal{T}_{CDW}$
and $\mathcal{T}_{SDW}$ stand for charge density wave and spin density wave correlations (the subscripts $+$ and $-$ refer to, respectively, the real and
imaginary parts up to a constant of the corresponding order parameter). The RG flow equations at two loops then become

\vspace{-0.4cm}

\begin{eqnarray}
&&\dot{\mathcal{T}}_{SC}^{(s^{\pm})}=\left(u_3 - u_4 - \frac{u_3^{2}}{2}\right) \mathcal{T}_{SC}^{(s^{\pm})}, \\
&&\dot{\mathcal{T}}_{SC}^{(s)}=-\left(u_3 + u_4 + \frac{u_3^{2}}{2}\right) \mathcal{T}_{SC}^{(s)}, \\
&&\dot{\mathcal{T}}_{CDW\pm}=\left(u_1 - 2u_2 \mp u_3 - \frac{u_3^{2}}{2}\right) \mathcal{T}_{CDW\pm},\\
&&\dot{\mathcal{T}}_{SDW\pm}=\left(u_1 \pm u_3 - \frac{u_3^{2}}{2}\right) \mathcal{T}_{SDW\pm},
\end{eqnarray}

\noindent with initial conditions given by $\mathcal{T}_{i}(\xi=0)=\mathcal{T}_{i}^{(0)}$
for all order parameters defined above.

\begin{figure}[t]
  \includegraphics[height=5.4cm]{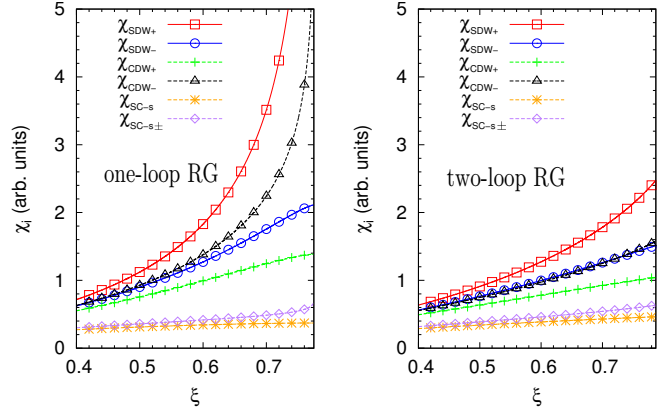}
  \caption{(Color online) RG flows of several susceptibilities of the model for both one-loop (left panel) and
  two-loop (right panel) RG approaches. The initial conditions used were $u_{1}^{(0)}=u_{4}^{(0)}=1$ and $u_{2}^{(0)}=u_{3}^{(0)}=0.1$.}\label{Diagrams}
\end{figure}

Once we computed
the response vertices associated with an instability towards a given ordered phase, we can then proceed to calculate their corresponding
susceptibilities. They are given by

\vspace{-0.2cm}

\begin{equation}\label{susc}
\chi_{i}(\xi)=\int_{0}^{\xi} d\zeta\mathcal{T}_{i}(\zeta)\mathcal{T}^{*}_{i}(\zeta),
\end{equation}

\noindent where $i=SC-s^{\pm},SC-s,CDW\pm$ and $SDW\pm$. In Fig. 6, we plot the results both for one-loop (which agrees with Ref. \cite{Chubukov2}) and the two-loop RG approach implemented here. We observe that the instabilities of the model within our RG scheme are surprisingly not changed qualitatively compared to one-loop results, even though the divergence at two loops takes place more slowly from a quantitative point of view.

\textbf{Conclusion.} -- We have carried out a complete two-loop RG study of an effective low-energy two-band model to describe some recently discovered iron-based superconductors. We have shown that the inclusion of two-loop quantum fluctuations has the main effect of changing the fixed point structure of the model. Even though the main instabilities turn out to be qualitatively the same as in one-loop calculations, the present work suggests a different microscopic mechanism with the dominant interactions at low energies being the intraband and interband forward interactions. It would be very interesting to implement such a two-loop RG scheme also at finite temperatures in order to calculate other physical quantities such as the uniform charge and spin susceptibilities of the model and to further compare the theoretical predictions obtained from the present RG approach with recent experimental studies \cite{Klingeler,XFWang} performed on these important materials.

\textbf{Acknowledgments.} -- We acknowledge support from the Brazilian agency CNPq through grant No. 474109/2010-0 for this project.

\end{document}